\documentclass[11pt]{article} 
\usepackage{epsfig}

\begin{document}
\input amssym.def 
\input amssym
\makeatletter
\setlength\@fptop{0\p@ }
\setlength\@fpsep{12\p@ }
\setlength\@fpbot{0\p@ plus 1fil }
\makeatother

\def\textfraction{.01}
\def\floatpagefraction{.8}

\intextsep 20pt plus 2pt minus 2pt
\setcounter{topnumber}{4}
\def\topfraction{.9}
\setcounter{bottomnumber}{2}
\def\bottomfraction{.8}
\setcounter{totalnumber}{6}

%
\makeatletter
\newinsert \@kludgeins
\global\dimen\@kludgeins \maxdimen
\global\count\@kludgeins 1000
\gdef \enlargethispage {%
   \@ifstar
     {%
      \@enlargepage{\hbox{\kern\p@}}}%
     {%
      \@enlargepage\@empty}%
}
\gdef\@enlargepage#1#2{%
   \@tempskipa#2\relax
   \ifdim \@tempskipa>.5\maxdimen
     \@latexerr{Suggested\space extra\space height\space
                (\the\@tempskipa)\space dangerously\space
                large}\@eha
   \else
     \ifdim \vsize<.5\maxdimen
       \@bsphack
         \insert\@kludgeins{#1\vskip-\@tempskipa}%
       \@esphack
     \else
       \@latexerr{Page\space height\space already\space
                  too\space large}\@eha
     \fi
   \fi
}
\makeatother

\hfuzz=5.0pt
%
%
%
%
\def\vec#1{\mathchoice{\mbox{\boldmath$\displaystyle\bf#1$}}
{\mbox{\boldmath$\textstyle\bf#1$}}
{\mbox{\boldmath$\scriptstyle\bf#1$}}
{\mbox{\boldmath$\scriptscriptstyle\bf#1$}}}
\def\mbf#1{{\mathchoice {\hbox{$\rm\textstyle #1$}}
{\hbox{$\rm\textstyle #1$}} {\hbox{$\rm\scriptstyle #1$}}
{\hbox{$\rm\scriptscriptstyle #1$}}}}
\def\operatorname#1{{\mathchoice{\rm #1}{\rm #1}{\rm #1}{\rm #1}}}
\chardef\ii="10
\def\widehat{\mathaccent"0362 }
\def\widetilde{\mathaccent"0365 }
\def\vphi{\varphi}
\def\vrho{\varrho}
\def\vtheta{\vartheta}
\def\ih{{\i\over\hbar}}
\def\hi{\frac{\hbar}{\i}}
\def\CD{{\cal D}}
\def\CE{{\cal E}}
\def\CH{{\cal H}}
\def\CL{{\cal L}}
\def\CP{{\cal P}}
\def\CV{{\cal V}}
\def\half{{1\over2}}
\def\bhalf{\hbox{$\half$}}
\def\viert{{1\over4}}
\def\halfvphi{\hbox{$\frac{\vphi}{2}$}}
\def\halfvphis{\hbox{$\frac{\vphi}{2}$}}
\def\halfvphiss{\hbox{$\frac{\vphi''}{2}$}}
\def\halfvtheta{\hbox{$\frac{\vtheta}{2}$}}
\def\halfvthetas{\hbox{$\frac{\vtheta'}{2}$}}
\def\halfvthetass{\hbox{$\frac{\vtheta''}{2}$}}
\def\bviert{\hbox{$\viert$}}
\def\hhbox#1#2{\hbox{$\frac{#1}{#2}$}}
\def\dfrac#1#2{\frac{\displaystyle #1}{\displaystyle #2}}
\def\intT{\ih\int_0^\infty\d\,T\,e^{\i ET/\hbar}}
\def\pathint#1{\int\limits_{#1(t')=#1'}^{#1(t'')=#1''}\CD #1(t)}
\def\hbarm{{\dfrac{\hbar^2}{2m}}}
\def\hbarmq{{\dfrac{\hbar^2}{2mq}}}
\def\mzwei{\dfrac{m}{2}}
\def\overh{\dfrac1\hbar}
\def\ihbar{\dfrac\i\hbar}
\def\intt{\int_{t'}^{t''}}
\def\tn{\tilde n}
\def\pmb#1{\setbox0=\hbox{#1}
    \kern-.025em\copy0\kern-\wd0
    \kern.05em\copy0\kern-\wd0
    \kern-.025em\raise.0433em\box0}
\def\pathintG#1#2{\int\limits_{#1(t')=#1'}^{#1(t'')=#1''}\CD_{#2}#1(t)}
\def\limN{\lim_{N\to\infty}}
\def\Norm{\bigg({m\over2\pi\i\epsilon\hbar}\bigg)}
\def\hbaram{{\hbar^2\over8m}}
\def\bbbr{{\rm I\!R}}                                
\def\bbbn{{\rm I\!N}}                                
\def\bbbz{{\mathchoice {\hbox{$\sf\textstyle Z\kern-0.4em Z$}}
{\hbox{$\sf\textstyle Z\kern-0.4em Z$}}
{\hbox{$\sf\scriptstyle Z\kern-0.3em Z$}}
{\hbox{$\sf\scriptscriptstyle Z\kern-0.2em Z$}}}}    
\def\bbbc{{\mathchoice {\setbox0=\hbox{\rm C}\hbox{\hbox
to0pt{\kern0.4\wd0\vrule height0.9\ht0\hss}\box0}}
{\setbox0=\hbox{$\textstyle\hbox{\rm C}$}\hbox{\hbox
to0pt{\kern0.4\wd0\vrule height0.9\ht0\hss}\box0}}
{\setbox0=\hbox{$\scriptstyle\hbox{\rm C}$}\hbox{\hbox
to0pt{\kern0.4\wd0\vrule height0.9\ht0\hss}\box0}}
{\setbox0=\hbox{$\scriptscriptstyle\hbox{\rm C}$}\hbox{\hbox
to0pt{\kern0.4\wd0\vrule height0.9\ht0\hss}\box0}}}}
\def\CP{{\cal P}}
\def\CQ{{\cal Q}}
\def\Ai{\operatorname{Ai}} 
\def\Cl{\operatorname{Cl}} 
\def\SU{\operatorname{SU}} 
\def\dt{\d t}
\def\d{\operatorname{d}}
\def\e{\operatorname{e}}
\def\i{\operatorname{i}}
\def\sn{\operatorname{sn}}
\def\cn{\operatorname{cn}}
\def\dn{\operatorname{dn}}
\def\max{\operatorname{max}}
\def\DI{D_{\,\rm I}}
\def\DII{D_{\,\rm II}}
\def\3dDII{D_{\,3d-\rm II}}
\def\DIII{D_{\,\rm III}}
\def\DIV{D_{\,\rm IV}}
\def\KI{K_{\,\rm I}}
\def\KInull{K_{\,\rm I}^{(0)}}
\def\KII{K_{\,\rm II}}
\def\KIII{K_{\,\rm III}}
\def\KIV{K_{\,\rm IV}}
\def\KV{K_{\,\rm V}}
\def\vphi{\varphi}
\def\halfphi{\hbox{$\frac{\vphi}{2}$}}
\def\tvphi{{\tilde\varphi}}
\def\tomega{{\tilde\omega}}
\def\ttau{{\tilde\tau}}
\def\hvphi{{\hat\varphi}}
\def\homega{{\hat\omega}}
\def\htau{{\hat\tau}}
\def\ps{\operatorname{ps}}
\def\Ps{\operatorname{Ps}}
\def\Si{\operatorname{Si}}
\def\energyldrei{\e^{-\i\hbar T(p^2+1)/2m}}
\def\ints{\int_0^{s''}}
\def\OO{\operatorname{O}}
\def\SO{\operatorname{SO}}
\def\fg{{\frak g}}
\def\fs{{\frak s}}
\def\fl{{\frak l}}
\def\gsl{\fg\fs\fl}
\def\operatorname#1{{\mathchoice{\rm #1}{\rm #1}{\rm #1}{\rm #1}}}
\def\bbbone{{\mathchoice {\rm 1\mskip-4mu l} {\rm 1\mskip-4mu l}
{\rm 1\mskip-4.5mu l} {\rm 1\mskip-5mu l}}}
\def\pathint#1{\int\limits_{#1(t')=#1'}^{#1(t'')=#1''}\CD #1(t)}
\def\pathints#1{\int\limits_{#1(0)=#1'}^{#1(s'')=#1''}\CD #1(s)}
 
\begin{titlepage}
\centerline{\normalsize DESY 07--132 \hfill ISSN 0418 - 9833}
\vskip.3in
\message{TITLE:}
\begin{center}
{\bf\Large Path Integral Approach for Quantum Motion 
\\[3mm]
on Spaces of Non-constant Curvature
\\[3mm]
According to Koenigs: Three Dimensions}
\end{center}
\message{Path Integral Approach for for Quantum Motion 
on Spaces of Non-constant Curvature According to Koenigs: Three Dimensions}
\vskip.5in
\begin{center}
{\large Christian Grosche}
\vskip.1in
{\normalsize\em II.\,Institut f\"ur Theoretische Physik}
\vskip.05in
{\normalsize\em Universit\"at Hamburg, Luruper Chaussee 149}
\vskip.05in
{\normalsize\em 22761 Hamburg, Germany}
\end{center}
\vskip2in
\normalsize

\vfill
\begin{center}
{\bf Abstract}
\end{center}
In this contribution a path integral approach for the quantum motion
on three-dimensional spaces according to Koenigs, for short``Koenigs-Spaces'',
is discussed.
Their construction is simple: One takes a Hamiltonian from three-dimensional
flat space and divides it by a three-dimensional superintegrable potential. 
Such superintegrable potentials will be the isotropic singular oscillator, the 
Holt-potential, the Coulomb potential, or two centrifugal potentials, respectively. 
In all cases a non-trivial space of non-constant curvature is generated. 
In order to obtain a proper quantum theory a curvature term has to be
incorporated into the quantum Hamiltonian. 
For possible bound-state solutions we find equations up to
twelfth order in the energy $E$. 
\end{titlepage}
\normalsize 
 
%

\setcounter{page}{1}%
\setcounter{equation}{0}%
\section{Introduction}%
\message{Introduction}%
In this contribution I discuss the quantum motion on three-dimensional 
spaces of non-constant curvature according to Koenigs \cite{KOENIGS}, 
which I will call for short ``Koenigs-spaces''. 
The construction of such a space is simple. One takes
a three-dimensional flat Hamiltonian, $\CH$, including some potential $V$, and
divides $\CH$ by a function $f(x,y,z)$ ($(x,y,z)\in\bbbr^3$) such that $f$
takes on the form of a metric:
\begin{equation}
\CH_{\rm Koenigs}=\frac{\CH}{f(x,y,z)}\enspace.
\end{equation}
Such a construction leads to a very rich structure, and attempts to classify
such systems are e.g. due to Kalnins et al. \cite{KKPM,KKMb} and 
Daskaloyannis and Ypsilantis \cite{DASYPS}.

\begin{table}[b!]
\caption{\label{tableE3} Coordinates in three-dimensional Euclidean space}
\hfuzz=15pt
\begin{eqnarray}\begin{array}{l}\vbox{\small\offinterlineskip
\halign{&\vrule#&$\strut\ \hfil\hbox{#}\hfill\ $\cr
\noalign{\hrule}
height2pt&\omit&&\omit&\cr
&Coordinate System &&Coordinates                                  &\cr
height2pt&\omit&&\omit&\cr
\noalign{\hrule}\noalign{\hrule}
height2pt&\omit&&\omit&\cr
&I.~Cartesian &&$x=x'$, $y=y'$,  $z=z'$                           &\cr
height2pt&\omit&&\omit&\cr
\noalign{\hrule}height2pt&\omit&&\omit&\cr
&II.~Circular Polar
              &&$x=\vrho\cos\vphi$, $y=\vrho\sin\vphi$, $z=z'$            &\cr
height2pt&\omit&&\omit&\cr
\noalign{\hrule}height2pt&\omit&&\omit&\cr
&III.~Circular Elliptic
              &&$x=d\cosh\mu\cos\nu$, $y=d\sinh\mu\sin\nu$,$z=z'$ &\cr
height2pt&\omit&&\omit&\cr
\noalign{\hrule}height2pt&\omit&&\omit&\cr
&IV.~Circular Parabolic
              &&$x=\half(\eta^2-\xi^2)$, $y=\xi\eta$, $z=z'$      &\cr
height2pt&\omit&&\omit&\cr
\noalign{\hrule}height2pt&\omit&&\omit&\cr
&V.~Sphero-Conical
              &&$x=r\sn(\alpha,k)\dn(\beta,k')$, 
                $y=r\cn(\alpha,k)\cn(\beta,k')$&\cr 
&             &&$z=r\dn(\alpha,k)\sn(\beta,k')$&\cr
height2pt&\omit&&\omit&\cr
\noalign{\hrule}height2pt&\omit&&\omit&\cr
&VI.~Spherical&&$x=r\sin\vtheta\cos\vphi$, 
                $y=r\sin\vtheta\sin\vphi$, 
                $z=r\cos\vtheta$               &\cr
height2pt&\omit&&\omit&\cr
\noalign{\hrule}height2pt&\omit&&\omit&\cr
&VII.~Parabolic&&$x=\xi\eta\cos\vphi$, 
                 $y=\xi\eta\sin\vphi$, 
                 $z=\half(\eta^2-\xi^2)$        &\cr
height2pt&\omit&&\omit&\cr
\noalign{\hrule}height2pt&\omit&&\omit&\cr
&VIII.~Prolate Spheroidal
              &&$x=d\sinh\mu\sin\nu\cos\vphi$, 
                $y=d\sinh\mu\sin\nu\sin\vphi$  &\cr
&             &&$z=d\cosh\mu\cos\nu$           &\cr
height2pt&\omit&&\omit&\cr
\noalign{\hrule}height2pt&\omit&&\omit&\cr
&IX.~Oblate Spheroidal
              &&$x=d\cosh\mu\sin\nu\sin\vphi$, 
                $y=d\cosh\mu\sin\nu\sin\vphi$  &\cr
&             &&$z=d\sinh\mu\cos\nu$           &\cr
height2pt&\omit&&\omit&\cr
\noalign{\hrule}height2pt&\omit&&\omit&\cr
&X.~Ellipsoidal&&$x=k^2\sqrt{a^2-c^2}\,\sn\alpha\sn\beta\sn\gamma$
                                                           &\cr
&   &&$y=-(k^2/k')\sqrt{a^2-c^2}\,\cn\alpha\cn\beta\cn\gamma$&\cr
&   &&$z=(\i/k')\sqrt{a^2-c^2}\,\dn\alpha\dn\beta\dn\gamma$  &\cr
height2pt&\omit&&\omit&\cr
\noalign{\hrule}height2pt&\omit&&\omit&\cr
&XI.~Paraboloidal&&$x=2d\cosh\alpha\cos\beta\sinh\gamma$, 
      $y=2d\sinh\alpha\sin\beta\cosh\gamma$                &\cr
&   &&$z=d(\cosh^2\alpha+\cos^2\beta-\cosh^2\gamma)$       &\cr
height2pt&\omit&&\omit&\cr
\noalign{\hrule}}}\end{array}\nonumber\end{eqnarray}
\vspace{-1cm}
\end{table}
\hfuzz=5pt
 
Simpler examples of such spaces are the two- and three-dimensional 
Darboux spaces, where one chooses the 
function $f$ in such a way that it depends only on one variable 
\cite{GROas,KalninsKMWinter}, respectively their
three-dimensional analogue \cite{GROat}. Another choice consists whether 
one chooses for $f$ some arbitrary potential (or some superintegrable
potential) and taking into account that the Poisson bracket structure of the 
observables makes up a reasonable simple algebra 
\cite{DASYPS,FMSUW,KalninsKMWinter}.

In previous publications we have analyzed the quantum motion on Darboux spaces
by means of the path integral \cite{GROas,GROPOe} and on two-dimensional
Koenigs-spaces \cite{GROau}. The path integral approach
\cite{FH,GRSh,KLEINERT,SCHULMAN} served as a powerful tool to calculate the 
propagator, respectively the Green function of the quantum motion in such
spaces. In the present contribution I apply the path integral technique to
five kinds of Koenigs-spaces, where a specific three-dimensional
superintegrable potential \cite{GROPOa} for the function $f$ is chosen. They are the
three-dimensional isotropic singular oscillator (Section II), 
the Holt-potential (section III), the three-dimensional Coulomb-potential
(Section IV), and two centrifugal potentials (Section V  and VI). 
The last Section is devoted to a summary and a discussion of the
achieved results.

\hfuzz=45pt
\begin{table}[t!]
\caption{\label{tableSWP3} The three-dimensional maximally
            super-integrable potentials\hfill}
\hfuzz=45pt
\begin{eqnarray}
\begin{array}{l}
\vbox{\offinterlineskip
\hrule
\halign{&\vrule#&
  \strut\quad\hfil#\quad\hfill\ \cr
height2pt&\omit&&\omit&\cr
&Potential $V(x,y,z)$, $\vec x=(x,y,z)\in\bbbr^3$
  &&Coordinate System                                       &\cr
height2pt&\omit&&\omit&&\omit&\cr
\noalign{\hrule}\noalign{\hrule}
height2pt&\omit&&\omit&&\omit&\cr
&$\displaystyle
  V_1={M\over2}\omega^2\vec x^2
    +\hbarm\Bigg({k_1^2-\viert\over x^2}
    +{k_2^2-\viert\over y^2}+{k_3^2-\viert\over z^2}\Bigg)$
  &&$\underline{\hbox{Cartesian}}$              &\cr
& &&$\underline{\hbox{Spherical}}$              &\cr
& &&$\underline{\hbox{Circular Polar}}$         &\cr
& &&Circular Elliptic                           &\cr
& &&Conical                                     &\cr
& &&Oblate Spheroidal                           &\cr
& &&Prolate Spheroidal                          &\cr
& &&Ellipsoidal                                 &\cr
height2pt&\omit&&\omit&\cr
\noalign{\hrule}
height2pt&\omit&&\omit&\cr
&$\displaystyle
  V_2={M\over2}\omega^2(x^2+y^2+4z^2)
     +\hbarm\Bigg({k_1^2-\viert\over x^2}+{k_2^2-\viert\over y^2}\Bigg)$
  &&$\underline{\hbox{Cartesian}}$              &\cr
& &&Parabolic                                   &\cr
& &&$\underline{\hbox{Circular Polar}}$     &\cr
& &&Circular Elliptic                           &\cr
height2pt&\omit&&\omit&\cr
\noalign{\hrule}
height2pt&\omit&&\omit&\cr
&$\displaystyle
  V_3=-{\alpha\over\sqrt{x^2+y^2+z^2}}
    +\hbarm\Bigg({k_1^2-\viert\over x^2}
    +{k_2^2-\viert\over y^2}\Bigg)$
  &&Conical                                     &\cr
& &&$\underline{\hbox{Spherical}}$              &\cr
& &&$\underline{\hbox{Parabolic}}$              &\cr
& &&{Prolate Spheroidal II}                 &\cr
height2pt&\omit&&\omit&\cr
\noalign{\hrule}
height2pt&\omit&&\omit&\cr
&$\displaystyle
  V_4=\hbarm\bigg(
  {k_1^2x\over y^2\sqrt{x^2+y^2}}+{k_2^2-\viert\over y^2}
  +{k_3^2-\viert\over z^2}\Bigg)$
  &&$\underline{\hbox{Spherical}}$              &\cr
&
  &&{Circular Elliptic II}                  &\cr
& &&$\underline{\hbox{Circular Parabolic}}$     &\cr
& 
  &&$\underline{\hbox{Circular Polar}}$     &\cr
height2pt&\omit&&\omit&\cr
\noalign{\hrule}
height2pt&\omit&&\omit&\cr
&$\displaystyle
  V_5=\hbarm\bigg(
  {k_1^2x\over y^2\sqrt{x^2+y^2}}+{k_2^2-\viert\over y^2}\Bigg)-k_3z$
  &&$\underline{\hbox{Circular Polar}}$     &\cr
&
  &&{Circular Elliptic II}                  &\cr
& &&$\underline{\hbox{Circular Parabolic}}$     &\cr
& 
  &&Parabolic                                   &\cr
height2pt&\omit&&\omit&\cr}\hrule}
\end{array}
         \nonumber
\end{eqnarray}
\end{table}
\hfuzz=7.0pt

In Table \ref{tableE3} I have displayed the 11 coordinate systems in
$\bbbr^3$. In a previous article \cite{GROPOa} we have discussed in much detail
the minimally and maximally superintegrable systems in $\bbbr^3$. There are
five  maximally superintegrable and seven minimally superintegrable
system. The maximally superintegrable potentials have the property that these
systems have five functionally independent integrals of motion (classical
mechanics), respectively five observables (quantum mechanics).
The minimally superintegrable instead have only four functionally
independent integrals of motion, respectively four
observables. In \cite{GROPOa} we have called these
superintegrable systems ``Smorodinsky-Winternitz potentials''.
In Table \ref{tableSWP3} I have indicated the coordinate systems in which
the  five maximally superintegrable systems in $\bbbr^3$ are separable.
The cases where an explicit path
integration is possible are $\underline{\hbox{underlined}}$.


\setcounter{equation}{0}%
\section{Koenigs-Space 
$\KI$ with Isotropic Singular Oscillator}
\message{Koenigs-Space KI with Isotropic Singular Oscillator}%
We start with the first example, where we take for the metric terms
\begin{eqnarray}
\d s^2&=& f_{I}(x,y,z)(\d x^2+ \d y^2+\d z^2)\enspace,\\
f_{I}(x,y,z)&=& \alpha
(x^2+y^2+z^2)+\frac{\beta_x}{x^2}+\frac{\beta_y}{y^2}+\frac{\beta_z}{z^2}+
\delta\enspace, 
\end{eqnarray}
and $\alpha,\beta_x,\beta_y,\beta_z,\delta$ are constants.
The classical Hamiltonian and Lagrangian in $\bbbr^3$ with the isotropic
singular oscillator as the superintegrable potential have the form:
\begin{eqnarray}
\CL&=&\frac{m}{2}\Big((\dot x^2+\dot y^2+\dot z^2)-\omega^2(x^2+y^2+z^2)\Big)
-\hbarm\bigg(\frac{k_x^2+\half}{x^2}+\frac{k_y^2+\half}{y^2}
+\frac{k_z^2+\half}{z^2}\bigg)\,,\qquad\\
\CH&=&\frac{p_x^2+p_y^2+p_z^2}{2m}+\frac{m}{2}\omega^2(x^2+y^2+z^2)
+\hbarm\bigg(\frac{k_x^2+\half}{x^2}+\frac{k_y^2+\half}{y^2}
+\frac{k_z^2+\half}{z^2}\bigg)\enspace.
\end{eqnarray}
(The specific choice of the constant $+\half$ has practical reasons which
will become clear in the sequel.)
Counting constants, there are nine independent constants:
$\alpha,\beta_{x,y,z},\delta$, and $\omega,k_{x,y,z}$. 
A tenth constant can be added by adding a further constant $\tilde\delta$ 
into the potential of the Hamiltonian. It will be omitted in the following.
The first Koenigs-space $\KI$ is constructed by considering
\begin{equation}
\CH_{\KI}=\frac{\CH}{f_{I}(x,y,z)}\enspace,
\end{equation}
hence for the Lagrangian (with potential)
\begin{eqnarray}
\CL_{\KI}&=&\frac{m}{2}f_{I}(x,y,z)(\dot x^2+\dot y^2+\dot z^2)
\nonumber\\  & &\quad
 -\frac{1}{f_{I}(x,y,z)}\Bigg[\frac{m}{2}\omega^2(x^2+y^2+z^2)
 +\hbarm\left(\frac{k_x^2+\half}{x^2}+\frac{k_y^2+\half}{y^2}\right)
+\frac{k_z^2+\half}{z^2}\Bigg]\enspace.\qquad
\end{eqnarray}
Setting the potential in square-brackets equal to zero yields the Lagrangian
for the free motion in $\KI$. With this information we can set up the path
integral in $\KI$ including a potential. Because the space is 
three-dimensional, the quantum potential $\propto\hbar^2$ does not have the
simple form as in the two-dimensional case \cite{GROau}. This is due to the
fact that two-dimensional spaces are conformally flat, and has the consequence
that in the path integral the additional quantum potential $\propto\hbar^2$
can be set to zero by choosing an appropriate lattice. This lattice 
corresponds to the product form path integral, i.e. we have for diagonal
metric $g_{ab}=f_a^2\delta_{ab}$ the quantum potential
\begin{equation}
\Delta V=\frac{\hbar^2(D-2)}{8m}
\sum_a\frac{(D-4)f_{a,a}^2+2f_af_{a,aa}}{f_a^4}\enspace.
\end{equation}
Obviously, $\Delta V=0$ for $D=2$. 
For our purposes we rewrite the metric term in the following way:
\begin{equation}
f_{I}(x,y,z)=
\frac{\displaystyle \alpha x^2(x^2+y^2+z^2)
+\beta_x+\frac{x^2\beta_y}{y^2}+\frac{x^2\beta_z}{z^2}+
\delta x^2}{x^2}\equiv\frac{h_x^2}{x^2}\enspace,
\label{h_for-DeltaV}
\end{equation}
and similarly in terms of $h_y$ and $h_z$. 
This gives for the $x=x_1$-part of $\Delta V$
\begin{eqnarray}
\Delta V_x&=&\Delta V_{1,x}+\Delta V_{2,x}
\\
\Delta V_{1,x}&=&\frac{\hbar^2}{8m}
\frac{2x^2h_xh_{x,xx}-2xh_xh_{x,x}-x^2h_xh_{x,x}^2}{h_x^4}\enspace,\qquad
\Delta V_{2,x}=\frac{3\hbar^2}{8mh_x^2}\enspace.
\end{eqnarray}
Repeating the procedure for the $y=x_2$- and $z=x_3$-coordinate we get
\begin{eqnarray}
\Delta V&=&\Delta V_1+\Delta V_2=\sum_{i=1}^3\Delta V_{1,x_i}+\Delta V_2
\\
\Delta V_2&=&\frac{3\hbar^2}{8m}
\bigg(\frac{1}{h_x^2}+\frac{1}{h_y^2}+\frac{1}{h_z^2}\bigg)
=\frac{3\hbar^2}{8mf_I}
\bigg(\frac{1}{x^2}+\frac{1}{y^2}+\frac{1}{z^2}\bigg)\enspace.
\end{eqnarray}
Note that if we choose for $h\equiv1$ that there is only one summand in the
last equation with $\Delta V_2=\Delta V={3\hbar^2}/{8m}$ (this is the
case for the three-dimensional hyperboloid). 

\begin{table}[t!]
\caption{\label{cosytab1} Some special cases for the space $\KI$}
\begin{eqnarray}\begin{array}{l}\vbox{\small\offinterlineskip
\halign{&\vrule#&$\strut\ \hfill\hbox{#}\hfill\ $\cr
\noalign{\hrule}
height2pt&\omit&&\omit&&\omit&\cr
&Metric
 &&Space      
 &&$\Delta V$                                                   &\cr
height2pt&\omit&&\omit&&\omit&\cr
\noalign{\hrule}\noalign{\hrule}
height2pt&\omit&&\omit&&\omit&\cr
&$f_I(x,y,z)$
 &&Koenigs space $\KI$     
 &&$\displaystyle\Delta V_1+\frac{3\hbar^2}{8m}
   \bigg(\frac{1}{h_x^2}+\frac{1}{h_y^2}+\frac{1}{h_z^2}\bigg)$ &\cr
height2pt&\omit&&\omit&&\omit&\cr
\noalign{\hrule}
height2pt&\omit&&\omit&&\omit&\cr
&$\displaystyle\frac{bu^2-a}{u^2}$        
  &&Three-dimensional Darboux Space $\DII$       
  &&$\displaystyle\Delta V_1+\frac{3\hbar^2}{8m(bu^2-a)}$       &\cr
height2pt&\omit&&\omit&&\omit&\cr
\noalign{\hrule}
height2pt&\omit&&\omit&&\omit&\cr
&$\displaystyle\frac{1}{u^2}$        
 &&Three-dimensional Hyperboloid   
 &&$\displaystyle\frac{3\hbar^2}{8m}$     &\cr
height2pt&\omit&&\omit&&\omit&\cr
\noalign{\hrule}
height2pt&\omit&&\omit&&\omit&\cr
&$1$       &&$\bbbr^3$   &&$0$             &\cr
height2pt&\omit&&\omit&&\omit&\cr
\noalign{\hrule}}}\end{array}\nonumber\end{eqnarray}
\end{table}

In Table \ref{cosytab1} I have displayed some special cases of $\KI$.
From \cite{GROas} we know that the free motion in the three-dimensional
Darboux space $\DII$ is separable in all eleven coordinate systems listed in 
Table \ref{tableE3}.

We now repeat our reasoning from \cite{GROat}: The part $\Delta V_1$  disturbs
a proper quantum treatment of the three-dimensional Koenigs space, and we 
{\it set up our quantum theory with an effective Lagrangian} 
\begin{equation}
\CL_{\KI}^{\rm eff}=\CL_{\KI}+\Delta V_1\enspace.
\end{equation}
Actually, our effective Lagrangian corresponds to the subtraction of a 
curvature term in $\CH$ \cite{KaMih}.
The canonical momentum operators are constructed by
\begin{equation}
p_{x_i}=\hi\bigg(\frac{\partial}{\partial {x_i}}
+\frac{\Gamma_i}{2}\bigg)\enspace,\qquad
\Gamma_i=\frac{\partial}{\partial {x_i}}\ln\sqrt{g}\enspace,
\end{equation}
with $x_1=x,x_2=y,,x_3=z$ and $g=\det(g_{ab})$, $(g_{ab})$ the metric tensor.
The Hamiltonian then has the form
\begin{eqnarray}
\CH_{\KI}^{\rm eff}&=&
    -\frac{\hbar^2}{2m}\Delta_{LB}
    +\frac{1}{f_{I}(x,y,z)}\Bigg[\frac{m}{2}\omega^2(x^2+y^2+z^2)
\nonumber\\   & & \qquad\qquad\qquad\qquad
    +\hbarm\left(\frac{k_x^2+\half}{x^2}+\frac{k_y^2+\half}{y^2}
    +\frac{k_z^2+\half}{z^2}\right)\Bigg]-\Delta V_1
         \\   &=&
\frac{1}{2m}\frac{1}{\sqrt{f_I}}(p_x^2+p_y^2+p_z^2)\frac{1}{\sqrt{f_I}}
    +\frac{1}{f_{I}}\Bigg[\frac{m}{2}\omega^2(x^2+y^2+z^2)
\nonumber\\   & &\qquad\qquad\qquad\qquad
    +\hbarm\left(\frac{k_x^2+\half}{x^2}+\frac{k_y^2+\half}{y^2}
    +\frac{k_z^2+\half}{z^2}\right)\Bigg]+\Delta V_2\enspace.
\end{eqnarray}
For the path integral in the product lattice definition we obtain
by means of a space-time transformation \cite{GRSh,KLEINERT} 
($\Delta V_2$ inserted)
\begin{eqnarray}
&&\!\!\!\!\!\!\!\!\!\!\!\!
K^{(\KI)}(x'',x',y'',y',z',z'';T)
=\pathint{x}\pathint{y}\pathint{z}f_{I}(x,y,z)
\nonumber\\  &&\!\!\!\!\!\!\!\!\!\!\!\!\quad\times
\exp\Bigg(\ih\intt\Bigg\{\frac{m}{2}f_{I}(x,y,z)(\dot x^2+\dot y^2+\dot z^2)
\nonumber\\  &&\!\!\!\!\!\!\!\!\!\!\!\!\qquad
 -\frac{1}{f_{I}(x,y,z)}\Bigg[\frac{m}{2}\omega^2(x^2+y^2+z^2)
 +\hbarm\Bigg(\frac{k_x^2-\viert}{x^2}
             +\frac{k_y^2-\viert}{y^2}+\frac{k_z^2-\viert}{z^2}\Bigg)\Bigg]
 \Bigg\}\dt\Bigg)\qquad\qquad
\\  &&\!\!\!\!\!\!\!\!\!\!\!\!
G^{(\KI)}(x'',x',y'',y',z',z'';E)=\ih(f_I'f_I'')^{-\viert}\int_0^\infty \d s'' 
K^{(\KI)}(x'',x',y'',y',z',z'';s'')
                   \e^{\i\delta\cdot Es''/\hbar}\,,\qquad
\label{Green-integration}
\end{eqnarray}
(note the change of constant to $-\viert$)
with the time-transformed path integral $K^{(\KI)}(s'')$ given by
($\widetilde\omega^2=\omega^2-2\alpha E/m$)
\begin{eqnarray}
&&
K^{(\KI)}(x'',x',y'',y',z',z'';s'')  
=\pathints{x}\pathints{y}\pathints{z}
\nonumber\\  && \quad\times
\exp\Bigg\{\ih\ints\Bigg[\frac{m}{2}
   \Big((\dot x^2+\dot y^2+\dot z^2)-\widetilde\omega^2(x^2+y^2+z^2)\Big)
\nonumber\\  &&\qquad
   -\hbarm\Bigg(\frac{k_x^2-2m\beta_x E/\hbar^2-\viert}{x^2}
               +\frac{k_y^2-2m\beta_y E/\hbar^2-\viert}{y^2}
               +\frac{k_z^2-2m\beta_z E/\hbar^2-\viert}{z^2}\Bigg)\Bigg]
   \d s''\Bigg\}\,.
\nonumber\\  &&
\label{Ks-KI}
\end{eqnarray}
The path integrals in the variables $x,y,z$ are path integrals
for the radial harmonic oscillator, however with energy-dependent
coefficients. We also see that the only effect of the constant $\delta$
consists of an additional phase in $s''$-integral 
which has consequences for the energy spectrum.

\subsection{Koenigs-Space $\KI$ with Isotropic Singular Oscillator 
in Polar Coordinates}
\message{Koenigs-Space KI with Isotropic Singular Oscillator in Polar
  Coordinates}%
We switch in the usual way to three-dimensional 
polar coordinates $(r,\vtheta,\vphi)$, and abbreviate 
$\tilde k_x^2=k_x^2-2m\beta_x E/\hbar^2$, 
$\tilde k_y^2=k_y^2-2m\beta_y E/\hbar^2$ and
$\tilde k_z^2=k_y^2-2m\beta_z E/\hbar^2$, respectively. 
In the variables $\vtheta,\vphi$ we obtain path integrals for the 
P\"oschl--Potential, and in the variable $r$ a 
radial path integral. The successive path integrations therefore
yield
\begin{eqnarray}
&&K^{(\KI)}(r'',r',\vtheta'',\vtheta'',\vphi'',\vphi';s'')
=
\sum_{n_\vphi}\Phi_{n_\vphi}^{(\tilde k_y,\tilde k_x)}(\vphi'')
               \Phi_{n_\vphi}^{(\tilde k_y,\tilde k_x)}(\vphi')
\sum_{n_\theta}\Phi_{n_\theta}^{(\tilde k_z,\lambda_1)}(\theta'')
               \Phi_{n_\theta}^{(\tilde k_z,\lambda_1)}(\theta')
\nonumber\\  &&\qquad\qquad\times
\frac{m\widetilde\omega\sqrt{r'r''}}{\i\hbar\sin\widetilde\omega s''}
\exp\Bigg[-\frac{m\widetilde\omega}{2\i\hbar}({r'}^2+{r''}^2)
       \cot\widetilde\omega s''\Bigg]
I_{\lambda_2}\Bigg(\frac{m\widetilde\omega r'r''}
                    {\i\hbar\sin\widetilde\omega s''}\Bigg)\enspace.
\end{eqnarray}
Here $\lambda_1=2n_\vphi+\tilde k_x+\tilde k_y+1$, 
$\lambda_2=2n_\vtheta+\tilde k_z+\lambda_1+1$, and the 
$\Phi_{n_\vphi}^{(\tilde k_y,\tilde k_x)}(\vphi)$ are the wave-functions for
the P\"oschl-Teller potential, which are given by \cite{BJb,DURb,FLMa,KLEMUS}
\begin{eqnarray}
  V^{(PT)}(x)&=&\hbarm\bigg(
  {\alpha^2-{1\over4}\over\sin^2x}+{\beta^2-{1\over4}\over\cos^2x}\bigg)
           \\  
  \Phi_n^{(\alpha,\beta)}(x)
  &=&\bigg[2(\alpha+\beta+2l+1)
  {l!\Gamma(\alpha+\beta+l+1)\over\Gamma(\alpha+l+1)\Gamma(\beta+l+1)}
  \bigg]^{1/2}
  \nonumber\\   &&\qquad\qquad\times
  (\sin x)^{\alpha+1/2}(\cos x)^{\beta+1/2}
  P_n^{(\alpha,\beta)}(\cos2x)\enspace.
\end{eqnarray}
The $P_n^{(\alpha,\beta)}(z)$ are Gegenbauer polynomials \cite{GRA} and
$I_\lambda(z)$ is the modified Bessel function \cite{GRA}. Performing the
$s''$-integration we obtain the Green function $G^{(\KI)}(E)$ \cite{GRA,GRSh}:
\begin{eqnarray}
&&G^{(\KI)}(r'',r',\vtheta'',\vtheta'',\vphi'',\vphi';E)
\nonumber\\  &&
=(f_I'f_I'')^{-\viert}
 \sum_{n_\vphi}\Phi_{n_\vphi}^{(\tilde k_y,\tilde k_x)}(\vphi'')
               \Phi_{n_\vphi}^{(\tilde k_y,\tilde k_x)}(\vphi')
\sum_{n_\theta}\Phi_{n_\theta}^{(\tilde k_z,\lambda_1)}(\theta'')
               \Phi_{n_\theta}^{(\tilde k_z,\lambda_1)}(\theta')
\nonumber\\  &&\qquad\qquad\times
\frac{\Gamma\big[\half(1+\lambda_2-\delta\cdot E/\hbar\widetilde\omega)\big]}
     {\hbar\widetilde\omega\sqrt{r'r''}\,\Gamma(1+\lambda)}
W_{\delta\cdot E/2\widetilde\omega,\lambda_2/2}
  \bigg(\frac{m\widetilde\omega}{\hbar}r_>^2\bigg)
M_{\delta\cdot E/2\widetilde\omega,\lambda_2/2}
  \bigg(\frac{m\widetilde\omega}{\hbar}r_<^2\bigg)\enspace.\qquad
\label{Green-KI}
\end{eqnarray}
$M_{\mu,\nu}(z)$ and $W_{\mu,\nu}(z)$ are Whittaker-functions \cite{GRA}, and 
$r_<,r_>$ is the smaller/larger of $r',r''$.
The poles of the $\Gamma$-function give the energy-levels of the bound states:
\begin{equation}
\bhalf(1+\lambda_2-\delta\cdot E/\hbar\widetilde\omega)=-n_r\enspace,
\end{equation}
which is equivalent to ($N=n_r+n_\vtheta+n_\vphi=0,1,2,\dots$):
\begin{eqnarray}
\delta\cdot E  &=&\hbar\widetilde\omega(2N+\tilde k_x+\tilde k_y+\tilde k_z+3)
\label{Energy-KI}
\\
&=&\hbar\sqrt{\omega^2-\frac{2\alpha}{m}E}
    \left(2N+\sqrt{k_x^2-\frac{2m\beta_x}{\hbar^2}E} \,
        +\sqrt{k_y^2-\frac{2m\beta_y}{\hbar^2}E}
        +\sqrt{k_z^2-\frac{2m\beta_z}{\hbar^2}E}\,+3\right)\enspace.
\nonumber
\end{eqnarray}
In general, this quantization condition is an equation of twelfth order 
in $E$.  Such an equation cannot be solved generally, however, we cam study
some special cases:
\begin{enumerate}
\item The case $k_1=k_2=K_3=\omega=0$:
\newline
\begin{equation}
E_N=-\frac{2\alpha\hbar^2}{m}\frac{(2N+3)^2}{(\delta+2\sqrt{\alpha\beta}\,)^2}
\enspace.
\end{equation}
For $\alpha<0$ this gives an infinite well-defined bound state spectrum.
For $\alpha>0$ the spectrum is negative infinite. Usually this means that
a particle will fall into the center and the wave-functions are {\it not} well 
defined. However, let us recall that the spectrum on the SU(1,1) hyperboloid
gives a positive continuous spectrum {\it and} a negative infinite discrete spectrum.
Hence, unphysical for real particles such a spectrum can be given a physical
meaning nevertheless: One has to re-interprete the motion on the hyperboloid
(space with curvature) by dimensional reduction to a potential problem
in flat space: In the case of the SU(1,1) hyperboloid the modified 
P\"oschl--Teller potential emerges and the negative infinite spectrum is
gets a cut yielding only {\it finite number} of well-defined bound-states
\cite{BJb}.
\item The case $k_1=k_2=k_3=\alpha=0$:
\newline
\begin{equation}
E_N= -\frac{\beta^\omega2}{2\delta^2}
\left(1\mp\sqrt{1-\frac{2\delta\hbar}{m\beta\omega}(2N+3)}\,\right)^2
\enspace.
\end{equation}
This gives for $\beta\not=0$ semi-bound states with positive real part
(\ref{EnergyNbeta}).
\item The case $k_1=k_2=k_3=\alpha=\beta=0,\delta>0$:
\newline
\begin{equation}
E_N=\frac{\hbar\omega}{\delta}(2N+3)\enspace.
\label{EnergyNbeta}
\end{equation}
\item The case $\beta_1=\beta_2=\beta_3=\alpha=0,\delta>0$:
\newline
\begin{equation}
E_N=\frac{\hbar\omega}{\delta}(2N+k_1+k_2+k_3+3)\enspace,
\end{equation}
\end{enumerate}
and we recover the flat space limit.

If we know the bound state energy $E_N$, we can determine the
wave-functions according to
\begin{equation}
\Psi_N^{(\KI)}(r,\theta,\vphi)=N_Nf_{I}^{-1/4}
\Phi_{n_\vphi}^{(\tilde k_y,\tilde k_x)}(\vphi)
\Phi_{n_\theta}^{(\tilde k_z,\lambda_1)}(\theta)
\Phi_{n_r}^{(RHO,\lambda)}(r)\enspace,
\end{equation}
with the normalization constant $N_N$ determined by evaluating the residuum 
in the Green function (\ref{Green-KI}), and the $\Phi_N^{(RHO,\lambda)}(r)$
are the wave-functions of the radial harmonic oscillator \cite{GRSh}:
\begin{equation}
\Psi_n^{(RHO,\lambda)}(r)
=\sqrt{\frac{2m}{\hbar}\frac{n!}{\Gamma(n+\lambda+1)}\,r}
  \bigg({m\omega\over\hbar}r\bigg)^{\lambda/2}
  \exp\bigg(-{m\omega\over2\hbar}r^2\bigg)
  L_n^{(\lambda)}\bigg({m\omega\over\hbar}r^2\bigg)\enspace.
\end{equation}
We can recover the flat space limit with $\alpha=\beta_{x_i}=0$ with the
correct spectrum $E_N=\hbar\omega(N+k_x+k_y+k_z+3)/\delta$.

\subsection[Koenigs-Space $\KI$ with Isotropic Singular Oscillator 
in Cartesian Coordinates]{Koenigs-Space $\KI$ with Isotropic Singular Oscillator 
in Cartesian\\ Coordinates} 
\message{Koenigs-Space KI with Isotropic Singular Oscillator in Cartesian
  Coordinates}%
Instead of switching to polar coordinates we keep the Cartesian system and
obtain 
\begin{eqnarray}
&&\!\!\!\!\!\!\!\!K^{(\KI)}(x'',x',y'',y',z',z'';s'')
\nonumber\\  &&\!\!\!\!\!\!\!\!
=\prod_{i=1}^3
\frac{m\widetilde\omega\sqrt{x_i'x_i''}}{\i\hbar\sin\widetilde\omega s''}
\exp\Bigg[-\frac{m\widetilde\omega}{2\i\hbar}({x_i'}^2+{x_i''}^2)
       \cot\widetilde\omega s''\Bigg]
I_{\tilde k_{x_i}}\Bigg(\frac{m\widetilde\omega x_i'x_i''}
                    {\i\hbar\sin\widetilde\omega s''}\Bigg)
\nonumber\\  &&\!\!\!\!\!\!\!\!
=\sum_{n_x}\Phi_{n_x}^{(RHO,\tilde k_x)}(x')\Phi_{n_x}^{(RHO,\tilde k_x)}(x'')
\sum_{n_y}\Phi_{n_y}^{(RHO,\tilde k_y)}(y')\Phi_{n_y}^{(RHO,\tilde k_y)}(y'')
\nonumber\\  &&\!\!\!\!\!\!\!\!\qquad\quad\times
\sum_{n_z}\Phi_{n_z}^{(RHO,\tilde k_z)}(z')\Phi_{n_z}^{(RHO,\tilde k_z)}(z'')
\exp\bigg\{\ih\bigg[\hbar\widetilde\omega(
      2N+\tilde k_x+\tilde k_y+\tilde k_z+3\bigg)\bigg]s''\Bigg\}
\,.\,\qquad
\end{eqnarray}
Performing the $s''$-integration yields for the energy-spectrum the same
result as before. The wave-functions are given by
\begin{equation}
\Psi_N^{(\KI)}(x,y,z)=N_Nf_{I}^{-1/4}
\Phi_{n_x}^{(RHO,\tilde k_x)}(x)
\Phi_{n_x}^{(RHO,\tilde k_y)}(y)
\Phi_{n_z}^{(RHO,\tilde k_z)}(z)\enspace,
\end{equation}
with the normalization constant $N_N$ determined by evaluating the residuum 
in the Green function for the energy-levels determined by (\ref{Energy-KI}). 
Note that all
coefficients $\tilde k_x,\tilde k_y,\tilde k_z$ are energy-dependent.  

As it is well-known \cite{GROPOa}, the singular isotropic is separable also
in circular polar, circular elliptic, conical, oblate and prolate spheroidal
and ellipsoidal coordinates, from which only the circular polar coordinate
system ($\vrho,\vphi,z$) allows an explicit solution which is very easily obtained:
The principal difference just consists of replacing the product of the
two radial oscillator wave-functions in $x$ and $y$ by a product of
a P\"oschl--Teller wave-function in $\vphi$ and radial oscillator
wave-function in $\vrho$ \cite{GROPOa}. The energy-spectrum, of course,
remains the same and is again determined by (\ref{Energy-KI}). We omit further
details because this case does not give anything new.

\subsection{Koenigs-Space $\KI$ with Zero Constants}
\message{Koenigs-Space without Potential}%
We now consider the Koenigs space $\KI$ with constants set to zero, 
denoted by $\KInull$. This gives for the corresponding space-time
transformed path integral (\ref{Ks-KI})
\begin{eqnarray}
&&
K^{(\KInull)}(x'',x',y'',y',z',z'';s'')  
=\pathints{x}\pathints{y}\pathints{z}
\nonumber\\  && \quad\times
\exp\Bigg\{\ih\ints\Bigg[\frac{m}{2}
   (\dot x^2+\dot y^2+\dot z^2)+\alpha E(x^2+y^2+z^2)
\nonumber\\  &&\qquad
   +\hbarm\Bigg(\frac{2m\beta_x E/\hbar^2+3/4}{x^2}
               +\frac{2m\beta_y E/\hbar^2+3/4}{y^2}
               +\frac{2m\beta_z E/\hbar^2+3/4}{z^2}\Bigg)\Bigg]
   \d s''\Bigg\}\,.\qquad\qquad
\label{Ks-KI0}
\end{eqnarray}
Obviously, this path integral can be separated in all the coordinate systems
in which the singular isotropic oscillator is separable.

Let us investigate the case for the Cartesian coordinates. We consider the
quantization condition (\ref{Energy-KI}). We have to set 
$\omega=k_x=k_y=k_z=0\alpha\not=0$, which yields:
\begin{equation}
\delta\cdot E_N 
=\hbar\sqrt{-\frac{2\alpha}{m}E_N}
    \left(2N+3+\tilde\beta\sqrt{-\frac{2m}{\hbar^2}E_N}\,\right)\enspace,
\label{Energy-KI0}
\end{equation}
with $\tilde\beta=\sqrt{\beta_x}+\sqrt{\beta_y}+\sqrt{\beta_z}$. This
quantization condition is a quadratic equation in the energy $E$ and 
has the solution
\begin{equation}
E_N=-\frac{8\alpha^2\tilde\beta^2}{m(\delta^2+4\tilde\beta^2}(2N+3)^2
\left(1\pm\sqrt{1-\frac{\hbar^2(\delta^2+4\tilde\beta^2)}
                       {4\alpha\tilde\beta^2}}\,\right)^2
\enspace.
\end{equation}
As an easy special case we consider $\tilde\beta=0$, then
\begin{equation}
E_N=\frac{2\alpha\hbar^2}{m\delta^2}(2N+3)^2\enspace.
\end{equation}
Therefore we obtain an infinite discrete spectrum for $\delta\not=0$. 
The spectrum can either be positive ($\alpha>0$), or negative ($\alpha<0$).
Such infinite negative energy spectra are well-known for spaces with
indefinite metric, for instance for the $\SU(1,1)$-manifold 
\cite{BJb,FLMa,KLEMUS}. Such spectra can be used by dimensional reduction
for potential problems in flat space yielding finite negative discrete spectra.
For the case of the infinite positive
spectrum we are done with the corresponding wave-functions:
\begin{equation}
\Psi_N^{(\KInull)}(n_x,n_y,n_z)=N_Nf_{I}^{-1/4}
\Phi_{n_x}^{(RHO,\tilde k_x)}(x)
\Phi_{n_x}^{(RHO,\tilde k_y)}(y)
\Phi_{n_z}^{(RHO,\tilde k_z)}(z)\enspace,
\end{equation}
with the normalization constant $N_N$ determined by evaluating the residuum 
in the Green function for the energy-levels (\ref{Energy-KI0}). Note that all
coefficients $\tilde k_x,\tilde k_y,\tilde k_z$ are energy-dependent. 
%
We omit the path integral
representations in the other coordinate systems.


\setcounter{equation}{0}%
\section{Koenigs-Space $\KII$ with Holt-Potential}
\message{Koenigs-Space KII with Holt-Potential}
Next we consider for the metric terms
\begin{eqnarray}
\d s^2&=& f_{II}(x,y,z)(\d x^2+ \d y^2+ \d z^2)\enspace,\\
f_{II}(x,y,z)&=&\alpha(x^2+y^2+4z^2)
  +\frac{\beta_x}{x^2}+\frac{\beta_y}{x^y}+\delta\enspace,
\end{eqnarray}
and $\alpha,\beta_x,\beta_y,\delta$ are constants.
The classical Hamiltonian and Lagrangian in $\bbbr^3$ with the 
Holt-potential as the superintegrable potential have the form:
\begin{eqnarray}
\CL&=&\frac{m}{2}\Big((\dot x^2+\dot y^2+\dot z^2)-\omega^2(x^2+y^2+4z^2)\Big)
 -\hbarm\bigg(\frac{k_x^2+\half}{x^2}+\frac{k_y^2+\half}{y^2}\bigg)\enspace,\\
\CH&=&\frac{p_x^2+p_y^2+p_z^2}{2m}+\frac{m}{2}\omega^2(x^2+y^2+4z^2)
+\hbarm\bigg(\frac{k_x^2+\half}{x^2}+\frac{k_y^2+\half}{y^2}\bigg)
\enspace.
\end{eqnarray}
Counting constants, there are seven independent constants:
$\alpha,\beta_x,\beta_y,\delta$, and $\omega,k_x,k_y$. An eighth constant can
be added by adding a further constant $\tilde\delta$ into the potential
of the Hamiltonian, which is omitted. The second Koenigs-space $\KII$ with
potential is now constructed by considering 
\begin{equation}
\CH_{\KII}=\frac{\CH}{f_{II}(x,y,z)}\enspace.
\end{equation}
From the discussion in the Section II it is obvious how to construct the
path integral on $\KII$. 
Again, we introduce the functions $h$ similar as in  (\ref{h_for-DeltaV}),
but, there is now a new feature. From the construction of $\Delta V_2$
according to (\ref{h_for-DeltaV}) we obtain also a $z$-dependent term
$\propto 3\hbar^2/8m h_z^2$. For the three-dimensional Holt-potential
there is, however, no such term $\propto 1/z^2$. In fact, the same situation
occurs also for the Coulomb potential, see the next Section. Such a term 
$\propto 3\hbar^2/8m h_z^2$ would not spoil the separability in the Cartesian
coordinate system and the two circular systems (polar and parabolic), but it
spoils the separability in parabolic coordinates.

\begin{table}[t!]
\caption{\label{cosytab2} Some special cases for the space $\KII$}
\begin{eqnarray}\begin{array}{l}\vbox{\small\offinterlineskip
\halign{&\vrule#&$\strut\ \hfill\hbox{#}\hfill\ $\cr
\noalign{\hrule}
height2pt&\omit&&\omit&&\omit&\cr
&Metric
 &&Space      
 &&$\Delta V$                                                   &\cr
height2pt&\omit&&\omit&&\omit&\cr
\noalign{\hrule}\noalign{\hrule}
height2pt&\omit&&\omit&&\omit&\cr
&$f_{II}(x,y,z)$
 &&Koenigs space $\KII$     
 &&$\displaystyle\Delta V_1+\frac{3\hbar^2}{8m}
   \bigg(\frac{1}{h_x^2}+\frac{1}{h_y^2}+\frac{1}{h_z^2}\bigg)$ &\cr
height2pt&\omit&&\omit&&\omit&\cr
\noalign{\hrule}
height2pt&\omit&&\omit&&\omit&\cr
&$\displaystyle\frac{bu^2-a}{u^2}$        
  &&Three-dimensional Darboux Space $\DII$       
  &&$\displaystyle\Delta V_1+\frac{3\hbar^2}{8m(bu^2-a)}$       &\cr
height2pt&\omit&&\omit&&\omit&\cr
\noalign{\hrule}
height2pt&\omit&&\omit&&\omit&\cr
&$\displaystyle\frac{1}{u^2}$        
 &&Three-dimensional Hyperboloid   
 &&$\displaystyle\frac{3\hbar^2}{8m}$     &\cr
height2pt&\omit&&\omit&&\omit&\cr
\noalign{\hrule}
height2pt&\omit&&\omit&&\omit&\cr
&$1$       &&$\bbbr^3$   &&$0$             &\cr
height2pt&\omit&&\omit&&\omit&\cr
\noalign{\hrule}}}\end{array}\nonumber\end{eqnarray}
\end{table}

Similarly, as in the previous Section we make the choice of symmetry
preservation and add the critical term $\propto 3\hbar^2/8m h_z^2$ into the
Lagrangian such that it is canceled after quantization, i.e.
\begin{equation}
\CL_{\KII}^{\rm eff}=\CL_{\KII}+\Delta V_1+\Delta V_2(z)\enspace.
\end{equation}

In Table \ref{cosytab2} I have listed some special cases of the Koenigs space
$\KII$. It is in fact the same, up to scaling, as for $\KI$. 

We proceed straightforward to the time-transformed
path integral $K^{(\KII)}(s'')$ which has the form
\begin{eqnarray}
&&\!\!\!\!\!\!\!\!
K^{(\KII)}(x'',x',y'',y',z',z';s'')=\pathints{x}\pathints{y}\pathints{z}
\nonumber\\  &&\!\!\!\!\!\!\!\!\qquad\times
\exp\Bigg\{\ih\ints\Bigg[\frac{m}{2}
   \Big((\dot x^2+\dot y^2+\dot z^2)-\widetilde\omega^2(x^2+y^2+4z^2)\Big)
\nonumber\\  &&\qquad\qquad\qquad\qquad\qquad\qquad\qquad\qquad
   -\hbarm\bigg(\frac{\tilde k_x^2-\viert}{x^2}
       +\frac{\tilde k_y^2-\viert}{y^2}\bigg)\Bigg]\d s''\Bigg\}\enspace.\qquad
\end{eqnarray}
Again, $\widetilde\omega^2=\omega^2-2\alpha E/m$, 
$\tilde k_{x_{1,2}}^2=k_{x_{1,2}}^2-2m\beta_{x_{1,2}} E/\hbar^2$.
We have in the variables $x,y$ a singular oscillator with frequency
$\widetilde\omega$, and in the variable $z$ 
an oscillator with frequency $2\widetilde\omega$.

Form \cite{GROPOa} we know that the Holt-potential is separable
in four coordinate systems: Cartesian, parabolic circular polar and circular
elliptic coordinates, respectively. Only in  Cartesian and circular polar
coordinates a closed solution is possible.
In Cartesian coordinates we take the respective solution as expanded into the
wave-functions and get
\begin{eqnarray}
&&\!\!\!\!\!\!\!\!
K^{(\KII)}(x'',x',y'',y',z',z';s'')
\nonumber\\  &&\!\!\!\!\!\!\!\!
=
\sum_{n_x}\Psi_{n_x}^{(RHO,\tilde k_x)}(x'')\Psi_{n_x}^{(RHO,\tilde k_x)*}(x')
\sum_{n_y}\Psi_{n_y}^{(RHO,\tilde k_y)}(x'')\Psi_{n_y}^{(RHO,\tilde k_y)*}(y')
\sum_{n_z}\Psi_{n_z}^{(HO)}(z'')\Psi_{n_z}^{(HO)*}(z')
\nonumber\\  &&\!\!\!\!\!\!\!\!\qquad\qquad\qquad\qquad\qquad\qquad\times
\e^{-\i s''\widetilde\omega(2n_x+2n_y+n_z+5/2)+\tilde k_x+\tilde k_y)}\enspace.
\end{eqnarray}
Here, the $\Psi_{n_z}^{(HO)}(z)$ denote the wave-functions of the 
harmonic oscillator with its Hermite polynomials.
Performing the $s''$-integration similarly as in (\ref{Green-integration})
we get the quantization condition ($N=2(n_x+n_y)+n_z$)
\begin{equation}
\delta\cdot E_N=\hbar \bigg(\omega^2-\frac{2\alpha}{m}E_N\bigg)^{1/2}
\left(N+\sqrt{k_x^2-\frac{2m\beta_x}{\hbar^2}E_N}\,
+\sqrt{k_y^2-\frac{2m\beta_y}{\hbar^2}E_N}\,+\hbox{$\frac{5}{2}$}\right)\,. 
\label{Energy-KII}
\end{equation}
In general, this is an equation of eighth order in $E_N$.
The solution in terms of the wave-functions then has the form
\begin{equation}
\Psi_N^{(\KII)}(x,y,z)=
N_Nf_{II}^{-1/4}\Psi_{n_x}^{(RHO,\tilde k_x)}(x)
\Psi_{n_y}^{(RHO,\tilde k_y)}(y)
\Psi_{n_z}^{(HO)}(z)\enspace,
\end{equation}
and the normalization constant $N_N$ is determined by the residuum
of the corresponding Green function at the energy $E_N$ from 
(\ref{Energy-KII}). The correct flat space limit with 
$\alpha=\beta_x=\beta_y=0$ is easily recovered
with spectrum $E_N=\hbar\omega (2N+\hbox{$\frac{5}{2}$}+k_x+k_y)$, and
similarly other special cases as in the previous Section.

The case of the circular polar coordinate system is very easily obtained.
The principal difference just consists of replacing the product of the
two radial oscillator wave-functions in $x$ and $y$ by a product of
a P\"oschl--Teller wave-function in $\vphi$ and radial oscillator
wave-function in $\vrho$ \cite{GROPOa}. The energy-spectrum, of course,
remains the same and is again determined by (\ref{Energy-KII}).

Similarly as in $\KI$, we can also consider the case of all constants set to zero
in the space $\KII$, denoted by $\KII^{(0)}$. The calculations are very 
similar to the previous section, yielding the quantization condition
($\tilde\beta=\sqrt{\beta_x}+\sqrt{\beta_y}$)
\begin{equation}
E_N(\delta^2+4\tilde\beta^2)-\frac{2\alpha\hbar^2}{m}
(N+\hbox{$\frac{5}{2}$})^2=
-\frac{4\alpha\tilde\beta}{m}(N+\hbox{$\frac{5}{2}$})\sqrt{-2mE_N}\enspace.
\end{equation} 
As an easy special case we consider $\tilde\beta=0$, then
\begin{equation}
E_N=\frac{2\alpha\hbar^2}{m\delta^2}(N+\hbox{$\frac{5}{2}$})^2\enspace,
\end{equation}
which yields for $\delta\not=0$ either a positive discrete spectrum
($\alpha>0$) or negative discrete spectrum ($\alpha<0$), these case have 
been already discussed in the previous section.

\setcounter{equation}{0}%
\section{Koenigs-Space $\KIII$ with Coulomb-Potential}
\message{Koenigs-Space KIII with Coulomb-Potential}
In the next example we consider a metric which corresponds to the
three-dimensional Coulomb potential ($r^2=x^2+y^2+z^2$)
\begin{eqnarray}
\d s^2&=& f_{III}(x,y,z)(\d x^2+ \d y^2+ \d z^2)\enspace,\\
f_{III}(x,y,z)&=&-\frac{\alpha_1}{\sqrt{x^2+y^2+z^2}}
   +\frac{\beta}{x^2}+\frac{\gamma}{y^2}+\delta
\end{eqnarray}
and $\alpha_1,\beta,\gamma,\delta$ are constants.
The classical Hamiltonian and Lagrangian in $\bbbr^3$ with the 
Coulomb potential as the superintegrable potential have the form:
\begin{eqnarray}
\CL&=&\frac{m}{2}(\dot x^2+\dot y^2+\dot z^2)
  +{\alpha_2\over r}-{\hbar^2\over2mr^2\sin^2\vtheta}
   \Bigg({k_1^2-\viert\over\cos^2\vphi}
        +{k_2^2-\viert\over\sin^2\vphi}\Bigg)
 \\
 \CH&=&\frac{p_x^2+p_y^2+p_z^2}{2m}
   -{\alpha_2\over r}+{\hbar^2\over2mr^2\sin^2\vtheta}
   \Bigg({k_1^2-\viert\over\cos^2\vphi}
        +{k_2^2-\viert\over\sin^2\vphi}\Bigg)\enspace.
\end{eqnarray}
Counting constants, there are seven independent constants:
$\alpha_1,\beta,\gamma,\delta$, and $\alpha_2,k_1,k_2$. 
An eight constants can be added by adding a further constant 
$\tilde\delta$ into the potential of the Hamiltonian, which is again omitted.
The third Koenigs-space $\KIII$ is constructed by considering
\begin{equation}
\CH_{\KIII}=\frac{\CH}{f_{III}(x,y,z)}\enspace.
\end{equation}

\begin{table}[t!]
\caption{\label{cosytab3} Some special cases for the space $\KIII$}
\begin{eqnarray}\begin{array}{l}\vbox{\small\offinterlineskip
\halign{&\vrule#&$\strut\ \hfill\hbox{#}\hfill\ $\cr
\noalign{\hrule}
height2pt&\omit&&\omit&&\omit&\cr
&Metric
 &&Space      
 &&$\Delta V$                                                   &\cr
height2pt&\omit&&\omit&&\omit&\cr
\noalign{\hrule}\noalign{\hrule}
height2pt&\omit&&\omit&&\omit&\cr
&$f_{III}(x,y,z)$
 &&Koenigs space $\KIII$     
 &&$\displaystyle\Delta V_1+\frac{3\hbar^2}{8m}
   \bigg(\frac{1}{h_x^2}+\frac{1}{h_y^2}+\frac{1}{h_z^2}\bigg)$ &\cr
height2pt&\omit&&\omit&&\omit&\cr
\noalign{\hrule}
height2pt&\omit&&\omit&&\omit&\cr
&$\frac{\alpha_1}{r}$
 &&Special Koenigs space $\KIII^{\alpha_1}$     
 &&$\displaystyle\Delta V_1+\frac{3\hbar^2}{8m}
   \bigg(\frac{1}{h_x^2}+\frac{1}{h_y^2}+\frac{1}{h_z^2}\bigg)$ &\cr
height2pt&\omit&&\omit&&\omit&\cr
\noalign{\hrule}
height2pt&\omit&&\omit&&\omit&\cr
&$\displaystyle\frac{bu^2-a}{u^2}$        
  &&Three-dimensional Darboux Space $\DII$       
  &&$\displaystyle\Delta V_1+\frac{3\hbar^2}{8m(bu^2-a)}$       &\cr
height2pt&\omit&&\omit&&\omit&\cr
\noalign{\hrule}
height2pt&\omit&&\omit&&\omit&\cr
&$\displaystyle\frac{1}{u^2}$        
 &&Three-dimensional Hyperboloid   
 &&$\displaystyle\frac{3\hbar^2}{8m}$     &\cr
height2pt&\omit&&\omit&&\omit&\cr
\noalign{\hrule}
height2pt&\omit&&\omit&&\omit&\cr
&$1$       &&$\bbbr^3$   &&$0$             &\cr
height2pt&\omit&&\omit&&\omit&\cr
\noalign{\hrule}}}\end{array}\nonumber\end{eqnarray}
\end{table}

In Table \ref{cosytab3} I have displayed some special cases for $\KIII$.
Again, the special cases are very similar as in the two previous cases,
except for only $\alpha_1\not=0$, and $\alpha_1,\delta\not=0$.

We proceed to the time-transformed path integral
$K^{(\KIII)}(s'')$ which has the form
\begin{eqnarray}
&&\!\!\!\!\!\!\!\!\!\!
K^{(\KIII)}(r'',r',\vtheta'',\vtheta',\vphi'',\vphi';s'')
=\pathints{r}\pathints{\vtheta}\pathints{\vphi}r^2\sin\vtheta
\nonumber\\  
&&\!\!\!\!\!\!\!\!\!\!\quad\times
\exp\Bigg\{\ih\ints\!\Bigg[\frac{m}{2}\Big(
\dot r^2\!+\!r^2(\dot\vtheta^2\!+\!\sin^2\vtheta\dot\vphi^2)\Big)
\!+\!{\tilde\alpha\over r}\!-\!{\hbar^2\over2mr^2\sin^2\vtheta}
\Bigg({\tilde k_1^2\!-\!\viert\over\cos^2\vphi}
\!+\!{\tilde k_2^2\!-\!\viert\over\sin^2\vphi}
\Bigg)\!\Bigg]\d s''\!\Bigg\}\,.
\nonumber\\   & &
\end{eqnarray}
Here, $\tilde k_1^2=k_1^2-2m\beta E/\hbar^2$, 
$\tilde k_2^2=k_2^2-2m\gamma E/\hbar^2$,
$\tilde\alpha=\alpha_2-\alpha_1 E$.
This path integral for the Coulomb potential has been discussed extensively in 
literature  and the solution in terms of the Green function
has been obtained by many authors, e.g.\cite{DURUKLE,GRSh,GROm,INOb,STEb}. 
We obtain for the Green function in polar coordinates
($\lambda_1=2n_\vphi\pm\tilde k_1\pm\tilde k_2+1$, $\lambda_2=l+\lambda_1+
\half$, $\kappa=\tilde\alpha\sqrt{-m/2\delta\cdot E}/\hbar$)
\begin{eqnarray}
&&\!\!\!\!\!\!\!\!\!\!
G^{(\KIII)}(r'',r',\vtheta'',\vtheta',\vphi'',\vphi';E)
  =(f_{III}'f_{III}'')^{-\viert}\sum_{n_\vphi=0}^\infty
  \Phi^{(\pm\tilde k_2,\pm\tilde k_1)}_n(\vphi'')
  \Phi^{(\pm\tilde k_2,\pm\tilde k_1)}_n(\vphi')
  \qquad \nonumber\\   & &\!\!\!\!\!\!\!\!\!\!\!\!\!\!\!\! \qquad\times
  \sum_{l=0}^\infty(l+\lambda_1+\bhalf){\Gamma(l+\lambda_1+1)\over l!}
  P_{\lambda_1+l}^{-\lambda_1}(\cos\vtheta'')
  P_{\lambda_1+l}^{-\lambda_1}(\cos\vtheta')
  \qquad \nonumber\\   & &\!\!\!\!\!\!\!\!\!\!\!\!\!\!\!\! \qquad\times
  {1\over r'r''}{1\over\hbar}\sqrt{-{m\over2E}}\,
  {\Gamma(\half+\lambda_2-\kappa)\over\Gamma(2\lambda_2+1)}
   W_{\kappa,\lambda_2}\bigg(\sqrt{-8mE}\,{r_>\over\hbar}\bigg)
   M_{\kappa,\lambda_2}\bigg(\sqrt{-8mE}\,{r_<\over\hbar}\bigg)\enspace.
\label{Green-KIII}
\end{eqnarray}
Bound states are determined by the poles of the Green function, respectively
by the poles of the $\Gamma$-function, i.e.
\begin{equation}
\half+\lambda_2-\kappa=-n_r\enspace,
\end{equation}
or more explicitly
\begin{equation}
2+2n_\vphi+l+n_r+\sqrt{k_1^2-\frac{2m\beta E_N}{\hbar^2}}
+\sqrt{k_2^2-\frac{2m\gamma E_N}{\hbar^2}}
-\frac{\alpha_2-\alpha_1E_N}{\hbar}\sqrt{-\frac{m}{2\delta\cdot E_N}}
=0\enspace.
\label{Energy-KIII}
\end{equation}
This is again an equation of twelfth order in the energy $E$.

We consider some special cases of (\ref{Energy-KIII}):
\begin{enumerate}
\item 
For $\beta=\gamma=\alpha_1=0$ we obtain the usual Coulomb potential
energy spectrum ($N=2+2n_\phi+l+n_r)$:
\begin{equation}
E_N=-\frac{m\alpha_2^2}{2\delta\hbar^2(N+k_1+k_2)^2}\enspace.
\end{equation}
\item 
For $k_1=k_2=\alpha_2=0$ we obtain the special space $\KIII^{\alpha_1}$    
($\tilde\beta=\sqrt\beta+\sqrt\gamma$):
\begin{equation}
E_N=-\frac{\hbar^2N^2}{2m(\tilde\beta+\half\alpha_1/\sqrt\delta)^2}
\enspace.
\end{equation}
\item 
For $k_1=k_2=0$ we obtain the special space $\KIII^{\alpha_1}$ with an
additional Coulomb potential   
($\tilde N^2=N^2+2m\alpha_2\hat\beta/(\sqrt\delta \hbar^2)$,
$\hat\beta=\tilde\beta+\alpha_1/(2\sqrt\delta)$):
\begin{equation}
E_N=-\frac{\hbar^2\tilde N^2}{4m\hat\beta^2}
\left(1\mp\sqrt{1-
\frac{4m^2\alpha_2^2\hat\beta^2}{\delta\hbar^4\tilde N^4}}\,\right)
\enspace.
\end{equation}
Note that for the upper-sign in the square-root term we get well-defined 
bound states for $N\to\infty$:
\begin{equation}
E_N=-\frac{m\alpha_2^2}{2\delta\hbar^2\tilde N^2}
\enspace,
\end{equation}
i.e. a Coulomb spectrum. However, note the complicated involvement of the various
constants, in particular the shift $N^2\to\tilde N^2$.
\end{enumerate}

In either case the wave-functions are given by
\begin{eqnarray}
&&\Psi_{n_r,l,n_phi}(r,\theta,\phi)=f_{III}^{-1/4}
\Phi^{(\pm\tilde k_2,\pm\tilde k_1)}_n(\vphi)
\sqrt{(l+\lambda_1+\bhalf){\Gamma(l+\lambda_1+1)\over l!}}
  P_{\lambda_1+l}^{-\lambda_1}(\cos\vtheta'')
        \nonumber\\   & &\qquad\times
  N_N{2\over(n+\lambda_1+\half)^2}
  \bigg[{2l!\over a^3(l+\lambda_2+\half)
   \Gamma(l+2\lambda_2+1)}\bigg]^{1/2}
  \bigg({2r\over a(l+\lambda_2+\half)}\bigg)^{\lambda_2}
         \nonumber\\   & & \qquad\times
  \exp\bigg(-{r\over a(l+\lambda_2+\half)}\bigg)
  L_l^{(2\lambda_2)}\bigg({2r\over a(l+\lambda_2+\half)}\bigg)\enspace,
\end{eqnarray}
provided the spectrum is bounded from below
and the additional normalization constant $N_N$ is determined 
by the poles of the Green function (\ref{Green-KIII}) at the 
energy-levels determined by (\ref{Energy-KIII}).

As it is well-known, the Coulomb potential is also separable in conical,
parabolic and and prolate spheroidal coordinates \cite{GROPOa}. 
In conical and prolate spheroidal coordinates no closed solutions in
terms of well-known higher transcendental functions can be found.
In parabolic coordinates, we have the same dependence in the variable
$\vphi$ as for polar coordinates, and in the variables $\xi$ and $\eta$
we get for the discrete spectrum a product of Laguerre polynomials and
exponentials, actually wave-functions very similar as in the polar variable 
$r$. The discrete spectrum, of course, remains the same. 
In \cite{GROPOa} this has been discussed in great detail and will not be 
repeated here.

The continuous spectrum is usually given in terms of $M$-Whittaker functions.
In the present case, this is quite an involved problem due to the
complicated structure of the indices. Both indices $\kappa$ and 
$\lambda_2$ are complex valued. This, in general leads to an energy 
spectrum $E_p>c$ with some constant $c>0$. For instance, in the case 
of the three-dimensional hyperboloid the constant is given by 
$c=\hbar^2/2m$, whereas in the case of the Coulomb potential in flat 
space $c=0$.

 
\setcounter{equation}{0}%
\section{Koenigs-Space $\KIV$ with Centrifugal Potential I}
\message{Koenigs-Space KIV with Centrifugal Potential I}
In the next example we consider a metric which corresponds to the
three-dimensional centrifugal potential
\begin{eqnarray}
\d s^2&=& f_{IV}(x,y,z)(\d x^2+ \d y^2+ \d z^2)\enspace,\\
f_{IV}(x,y)&=&
   \frac{\hbar^2}{2m}\left(
   \frac{\alpha x}{y^2\sqrt{x^2+y^2}}
   +\frac{\beta}{y^2}+\frac{\gamma}{z^2}\right)+\delta
\end{eqnarray}
and $\alpha_1,\beta,\gamma,\delta$ are constants.
The classical Hamiltonian and Lagrangian in $\bbbr^3$ with 
this potential in spherical coordinates have the form
\begin{eqnarray}
\CL&=&\frac{m}{2}(\dot x^2+\dot y^2+\dot z^2)
   -\frac{\hbar^2}{2mr^2}\left(\frac{1}{\sin^2\vtheta}
   \bigg(\frac{k_1^2+k_2^2-\viert}{4\sin^2\halfphi}
   +\frac{k_2^2-k_1^2-\viert}{4\cos^2\halfphi}\bigg)
   +\frac{k_3^2-\viert}{\cos^2\vtheta}\right)
\enspace,\qquad\\
\CH&=&\frac{p_x^2+p_y^2+p_z^2}{2m}
  +\frac{\hbar^2}{2mr^2}\left(\frac{1}{\sin^2\vtheta}
   \bigg(\frac{k_1^2+k_2^2-\viert}{4\sin^2\halfphi}
   +\frac{k_2^2-k_1^2-\viert}{4\cos^2\halfphi}\bigg)
   +\frac{k_3^2-\viert}{\cos^2\vtheta}\right)\enspace.
\end{eqnarray}
Counting constants, there are seven independent constants:
$\alpha,\beta,\gamma,\delta$, and $k_1,k_2,k_3$. 
An eight constants can be added by adding a further constant 
$\tilde\delta$ into the potential of the Hamiltonian, which is omitted.

\begin{table}[t!]
\caption{\label{cosytab6} Some special cases for the space $\KIV$}
\begin{eqnarray}\begin{array}{l}\vbox{\small\offinterlineskip
\halign{&\vrule#&$\strut\ \hfill\hbox{#}\hfill\ $\cr
\noalign{\hrule}
height2pt&\omit&&\omit&&\omit&\cr
&Metric
 &&Space      
 &&$\Delta V$                                                   &\cr
height2pt&\omit&&\omit&&\omit&\cr
\noalign{\hrule}\noalign{\hrule}
height2pt&\omit&&\omit&&\omit&\cr
&$f_{IV}(x,y,z)$
 &&Koenigs space $\KIV$     
 &&$\displaystyle\Delta V_1+\frac{3\hbar^2}{8m}
   \bigg(\frac{1}{h_x^2}+\frac{1}{h_y^2}+\frac{1}{h_z^2}\bigg)$ &\cr
height2pt&\omit&&\omit&&\omit&\cr
\noalign{\hrule}
height2pt&\omit&&\omit&&\omit&\cr
&$\displaystyle\frac{bu^2-a}{u^2}$        
  &&Three-dimensional Darboux Space $\DII$       
  &&$\displaystyle\Delta V_1+\frac{3\hbar^2}{8m(bu^2-a)}$       &\cr
height2pt&\omit&&\omit&&\omit&\cr
\noalign{\hrule}
height2pt&\omit&&\omit&&\omit&\cr
&$\displaystyle\frac{1}{u^2}$        
 &&Three-dimensional Hyperboloid   
 &&$\displaystyle\frac{3\hbar^2}{8m}$     &\cr
height2pt&\omit&&\omit&&\omit&\cr
\noalign{\hrule}
height2pt&\omit&&\omit&&\omit&\cr
&$1$       &&$\bbbr^3$   &&$0$             &\cr
height2pt&\omit&&\omit&&\omit&\cr
\noalign{\hrule}}}\end{array}\nonumber\end{eqnarray}
\end{table}

In Table \ref{cosytab6} I have displayed some special cases for the space $\KIV$.
Again, the special cases are very similar as in the previous cases.

The fourth Koenigs-space $\KIV$ is constructed by considering
\begin{equation}
\CH_{\KIV}=\frac{\CH}{f_{IV}(x,y,z)}\enspace.
\end{equation}
We write down the path integral formulation for $K^{(\KIV)}(s'')$
\begin{eqnarray}
&&\!\!\!\!\!\!\!\!\!\!
K^{(\KIV)}(r'',r',\vtheta'',\vtheta',\vphi'',\vphi';s'')
=\pathints{r}\pathints{\vtheta}\pathints{\vphi}r^2\sin\vtheta
\nonumber\\  &&\!\!\!\!\!\!\!\!\!\!\quad\times
\exp\Bigg\{\ih\ints\Bigg[\frac{m}{2}\Big(
\dot r^2+r^2(\dot\vtheta^2+\sin^2\vtheta\dot\vphi^2)\Big)
\nonumber\\  &&\qquad\qquad\qquad
   -\frac{\hbar^2}{2mr^2}\left(\frac{1}{\sin^2\vtheta}
   \bigg(\frac{\tilde k_1^2+\tilde k_2^2-\viert}{4\sin^2\halfphi}
   +\frac{\tilde k_2^2-\tilde k_1^2-\viert}{4\cos^2\halfphi}\bigg)
   +\frac{\tilde k_3^2-\viert}{\cos^2\vtheta}\right)\Bigg]
   \d s''\!\Bigg\}\,.\qquad
\end{eqnarray}
Here, $\tilde k_1^2=k_2^2+k_1^2-2m(\beta+\alpha) E/\hbar^2$, 
      $\tilde k_2^2=k_2^2-k_1^2+2m(\beta-\alpha) E/\hbar^2$,
$\tilde k_3^2=k_3^2-2m\gamma E/\hbar$.
We obtain for the path integral $K^{(\KIV)}(s'')$
($\lambda_1=n+(\tilde k_1+\tilde k_2+1)/2$, $\lambda_2=
2m+\lambda_1\pm \tilde k_3+1$)
\begin{eqnarray}
&&\!\!\!\!\!\!\!\!\!\!
K^{(\KIV)}(r'',r',\vtheta'',\vtheta',\vphi'',\vphi';s'')
         \nonumber\\   & &\!\!\!\!\!\!\!\!
  =\frac{(r'r''\sin\theta'\sin\theta'')^{-1/2}}{2}
   \sum_{n=0}^\infty
   \Psi^{(\tilde k_1,\tilde k_2)}_n\bigg({\vphi'\over2}\bigg)
   \Psi^{(\tilde k_1,\tilde k_2)}_n\bigg({\vphi''\over2}\bigg)
   \sum_{m=0}^\infty \Phi^{(\lambda_1,\pm \tilde k_3)}_m(\vtheta')
  \Phi^{(\lambda_1,\pm \tilde k_3)}_m(\vtheta'')
         \nonumber\\   & &\!\!\!\!\!\!\!\! \qquad\times
  {m\over\i\hbar s''}\exp\bigg[{\i m\over2\hbar s''}({r'}^2+{r''}^2)\bigg]
  I_{\lambda_2}\bigg({mr'r''\over\i\hbar s''}\bigg)\enspace.
\end{eqnarray}
Therefore the Green function $G^{(\KIV)}(E)$ has the form
($\tilde E = E-\delta$)
\begin{eqnarray}
&&\!\!\!\!\!\!\!\!\!\!
G^{(\KIV)}(r'',r',\vtheta'',\vtheta',\vphi'',\vphi';E)
  =(f_{IV}'f_{IV}'')^{-\viert}\frac{(r'r''\sin\theta'\sin\theta'')^{-1/2}}{2}
         \nonumber\\   & &\!\!\!\!\!\!\!\! \qquad\times
   \sum_{n=0}^\infty
   \Psi^{(\tilde k_1,\tilde k_2)}_n\bigg({\vphi'\over2}\bigg)
   \Psi^{(\tilde k_1,\tilde k_2)}_n\bigg({\vphi''\over2}\bigg)
   \sum_{m=0}^\infty \Phi^{(\lambda_1,\pm \tilde k_3)}_m(\vtheta')
  \Phi^{(\lambda_1,\pm \tilde k_3)}_m(\vtheta'')
         \nonumber\\   & &\!\!\!\!\!\!\!\! \qquad\times
  \frac{2m}{\hbar^2}
  I_{\lambda_2}\bigg(\sqrt{-2m\tilde E}\,\frac{r_<}{\hbar}\bigg)
  K_{\lambda_2}\bigg(\sqrt{-2m\tilde E}\,\frac{r_>}{\hbar}\bigg)
\enspace.
\end{eqnarray}
The analysis of this Green function is complicated due to the 
complicated index which has imaginary parts for 
$2m\gamma E/\hbar^2>k_3^2$ etc. This, in general leads to 
modified K-Bessel-functions as wave-functions (c.f. Liouville quantum
mechanics \cite{GRSh}) with a continuous energy spectrum $E_p>c$ with some
constant $c>0$. For instance, in the case of the three-dimensional hyperboloid
the constant is given by $c=\hbar^2/2m$. We do not discuss these issues
any further.

 
\setcounter{equation}{0}%
\section{Koenigs-Space $\KV$ with Centrifugal Potential II}
\message{Koenigs-Space KV with Centrifugal Potential II}
In the last example we consider a metric which corresponds to the
three-dimensional linear-centrifugal potential
\begin{eqnarray}
\d s^2&=& f_{V}(x,y,z)(\d x^2+ \d y^2+ \d z^2)\enspace,\\
f_{V}(x,y)&=&
   \frac{\hbar^2}{2m}\left(\frac{\alpha x}{y^2\sqrt{x^2+y^2}}
   +\frac{\beta}{y^2}\right)+\gamma z+\delta
\end{eqnarray}
and $\alpha_1,\beta,\gamma,\delta$ are constants.
The classical Hamiltonian and Lagrangian in $\bbbr^3$ with 
this potential have the form
\begin{eqnarray}
\CL&=&\frac{m}{2}(\dot x^2+\dot y^2+\dot z^2)
   -\frac{\hbar^2}{2m}\left(\frac{k_1^2 x}{y^2\sqrt{x^2+y^2}}
   +\frac{k_2^2-\viert}{y^2}\right)-k_3 z
\enspace,\\
\CH&=&\frac{p_x^2+p_y^2+p_z^2}{2m}
   +\frac{\hbar^2}{2m}\left(\frac{k_1^2 x}{y^2\sqrt{x^2+y^2}}
   +\frac{k_2^2-\viert}{y^2}\right)+k_3 z\enspace.
\end{eqnarray}
Counting constants, there are seven independent constants:
$\alpha,\beta,\gamma,\delta$, and $k_1,k_2,k_3$. 
An eight constants can be added by adding a further constant 
$\tilde\delta$ into the potential of the Hamiltonian, which is omitted.
The fifth Koenigs-space $\KV$ is constructed by considering
\begin{equation}
\CH_{\KV}=\frac{\CH}{f_{V}(x,y,z)}\enspace.
\end{equation}

\begin{table}[t!]
\caption{\label{cosytab5} Some special cases for the space $\KV$}
\begin{eqnarray}\begin{array}{l}\vbox{\small\offinterlineskip
\halign{&\vrule#&$\strut\ \hfill\hbox{#}\hfill\ $\cr
\noalign{\hrule}
height2pt&\omit&&\omit&&\omit&\cr
&Metric
 &&Space      
 &&$\Delta V$                                                   &\cr
height2pt&\omit&&\omit&&\omit&\cr
\noalign{\hrule}\noalign{\hrule}
height2pt&\omit&&\omit&&\omit&\cr
&$f_{V}(x,y,z)$
 &&Koenigs space $\KV$     
 &&$\displaystyle\Delta V_1+\frac{3\hbar^2}{8m}
   \bigg(\frac{1}{h_x^2}+\frac{1}{h_y^2}+\frac{1}{h_z^2}\bigg)$ &\cr
height2pt&\omit&&\omit&&\omit&\cr
\noalign{\hrule}
height2pt&\omit&&\omit&&\omit&\cr
&$\gamma u$
 &&Three-dimensional Darboux space $\DI$    
 &&$0$ &\cr
height2pt&\omit&&\omit&&\omit&\cr
\noalign{\hrule}
height2pt&\omit&&\omit&&\omit&\cr
&$\displaystyle\frac{bu^2-a}{u^2}$        
  &&Three-dimensional Darboux Space $\DII$       
  &&$\displaystyle\Delta V_1+\frac{3\hbar^2}{8m(bu^2-a)}$       &\cr
height2pt&\omit&&\omit&&\omit&\cr
\noalign{\hrule}
height2pt&\omit&&\omit&&\omit&\cr
&$\displaystyle\frac{1}{u^2}$        
 &&Three-dimensional Hyperboloid   
 &&$\displaystyle\frac{3\hbar^2}{8m}$     &\cr
height2pt&\omit&&\omit&&\omit&\cr
\noalign{\hrule}
height2pt&\omit&&\omit&&\omit&\cr
&$1$       &&$\bbbr^3$   &&$0$             &\cr
height2pt&\omit&&\omit&&\omit&\cr
\noalign{\hrule}}}\end{array}\nonumber\end{eqnarray}
\end{table}

In Table \ref{cosytab5} I have displayed some special cases for $\KV$.
Again, the special cases are very similar as in the previous cases.
However, the new special case which appears is the three dimensional 
Darboux space $\DI$ from \cite{GROat}. From \cite{GROat} we know that
the free motion separates in seven coordinate systems, i.e. in
Cartesian, the three circular systems, the parabolic, the paraboloidal system,
and a rotated Cartesian system.

We write down the path integral formulation for $K^{(\KV)}(s'')$
in circular polar coordinates
\begin{eqnarray}
&&\!\!\!\!\!\!\!\!\!\!
K^{(\KV)}(\vrho'',\vrho',\vphi'',\vphi',z'',z';s'')
=\pathints{\vrho}\vrho\pathints{\vphi}\pathints{z}
\nonumber\\  &&\!\!\!\!\!\!\!\!\!\!\quad\times
\exp\left\{\ih\!\ints\!\left[\frac{m}{2}\Big(
\dot\vrho^2\!+\!\vrho^2\dot\vphi^2\!+\!\dot z^2)\Big)
\!-\!\frac{\hbar^2}{2m\vrho^2}
\left(\frac{\tilde k_1^2-\viert}{4\cos^2\halfphi}
   \!+\!\frac{\tilde k_2^2-\viert}{4\sin^2\halfphi}
   -\viert\right)-\!\tilde k_3 z\right]
   \d s''\!\right\}\,.\qquad
\end{eqnarray}
Here, $\tilde k_1^2=k_2^2+k_1^2-2m(\beta+\alpha) E/\hbar^2$, 
      $\tilde k_2^2=k_2^2-k_1^2+2m(\beta-\alpha) E/\hbar^2$,
$\tilde k_3=k_3-2m\gamma E/\hbar$.
This path integral has the solution
($\lambda_1=n+\half(\tilde k_1+\tilde k_2+1)$):
\begin{eqnarray}
         \nonumber\\   & &\!\!\!\!\!\!\!\!\!\!
K^{(\KV)}(\vrho'',\vrho',\vphi'',\vphi',z'',z';s'')
         \nonumber\\   & &\!\!\!\!\!\!\!\!\!\!
  =\bigg({m\over2\pi\i\hbar s''}\bigg)^{1/2}
  \exp\Bigg[\ih\Bigg({m\over2s''}(z''-z')^2-{k_3T\over2}(z'+z'')
      -{k_3^2s''^3\over24m}\Bigg)\Bigg]
         \nonumber\\   & &\!\!\!\!\!\!\!\!\!\!\qquad\times
  {m\over2\i\hbar s''}\sum_{n=0}^\infty
   \Phi_n^{(\tilde k_1,\tilde k_2)}\bigg({\phi'\over2}\bigg)
   \Phi_n^{(\tilde k_1,\tilde k_2)}\bigg({\phi''\over2}\bigg)
  \exp\bigg[-{m\over2\i\hbar s''}({\rho'}^2+{\rho''}^2)\bigg]
  I_{\lambda_1}\bigg({m\rho'\rho''\over\i\hbar s''}\bigg)
\,.\qquad
\end{eqnarray}
The analysis of this Green function is again very complicated due to the 
complicated index which has imaginary parts for 
$2m\gamma E/\hbar^2>k_3^2$ etc. yielding a continuous spectrum.
As in the previous section, these issues will not be discussed further.

 
\setcounter{equation}{0}%
\section{Summary and Discussion}
\message{Summary and Discussion}
In this contribution I have discussed a path integral approach for spaces of
non-constant curvature according to Koenigs, which I have for short called
``Koenigs-spaces'' $\KI$--$\KV$, respectively.
I have found a very rich structure of the spectral properties of the quantum
motion on Koenigs-spaces. 
In the general case with potential, in three spaces the
quantization condition is determined by an equation up to twelfth order in the
energy $E$. Such an equation cannot be solved explicitly, however special
cases can be studied. We found also constraints on the parameters for
the well-definedness on the wave-functions.
For the remaining two spaces no quantization condition was formulated,
because it is known that in the corresponding cases of 
superintegrable potentials in $\bbbr^3$ only a continuous spectrum exists.

Let us note a further feature of these spaces. It is obvious that our solutions
remain on a formal level. Neither have we specified an embedding
space, nor have we specified boundary conditions on our spaces.
Let us consider the space $\KV$: We set $\alpha=\beta=\delta =0$ and
$\gamma=1$. In this case we obtain a metric which corresponds to the 
three-dimensional Darboux space $\DI$ (modulo change of variables), 
as discussed in \cite{GROat}. In $\DI$ boundary 
conditions and the signature of the ambient space is very important, because
choosing a positive or a negative signature of the ambient space changes
the boundary conditions, and hence the quantization conditions 
\cite{GROPOe,KalninsKMWinter}. Including an appropriate potential,
bound states defined by a transcendental equation can be found.

Furthermore, we can recover the three-dimensional Darboux space $\DII$
\cite{GROat,KalninsKMWinter} by setting in our examples in the
potential function $f$ all constant to zero except those corresponding 
to the $1/x^2$-singularity and the constant $\delta=1$. 
However, we did not discuss these cases in detail.

In our approach we have chosen examples of superintegrable potentials
in three-di\-men\-sional space, i.e. the isotropic singular oscillator, 
the Holt potential, the Coulomb potential, and two centrifugal potentials, 
respectively. I did not consider the minimally superintegrable potentials.
There are eight of them \cite{GROPOa}, however they contain always an
unspecified function $F$ depending on the radial variable $r$, say, 
leading to some unspecified spectrum.

I have omitted the discussion of the continuous
spectrum. This is mostly due to lack of the specification of 
the ambient space. For instance, in the
Darboux space $\DII$ we know that the continuous spectrum
has the form of $E_p\propto(\hbar^2/2m)p^2+\hbox{constant}$. The
wave-functions are proportional to K-Bessel functions \cite{GROas}.
However, in Darboux space $\DI$ there is no such constant, and the
wave-functions have a different form. Furthermore, $\DII$ contains as special
cases  the three-dimensional Euclidean plane and the Hyperbolic space,
respectively. 

However, a more detailed study of these special case would
require some additional input from a physics point of view: Can a space
of non-constant curvature (Koenigs or Darboux space) model
actually curved space-time? And how such a global or local model
can give rise to observable physical effects? 
These issues are beyond the scope of this article and will not be
discussed here any further.

\subsection*{\bf Acknowledgments}
This work was supported by the Heisenberg--Landau program. 
I would like to thank G.Pogosyan, for the warm hospitality during 
my stay in Yerevan, Armenia. 
The author is grateful to Ernie Kalnins for fruitful and pleasant
discussions on superintegrability and separating coordinate systems.


\vbox{\centerline{\ }
\centerline{\epsfig{file=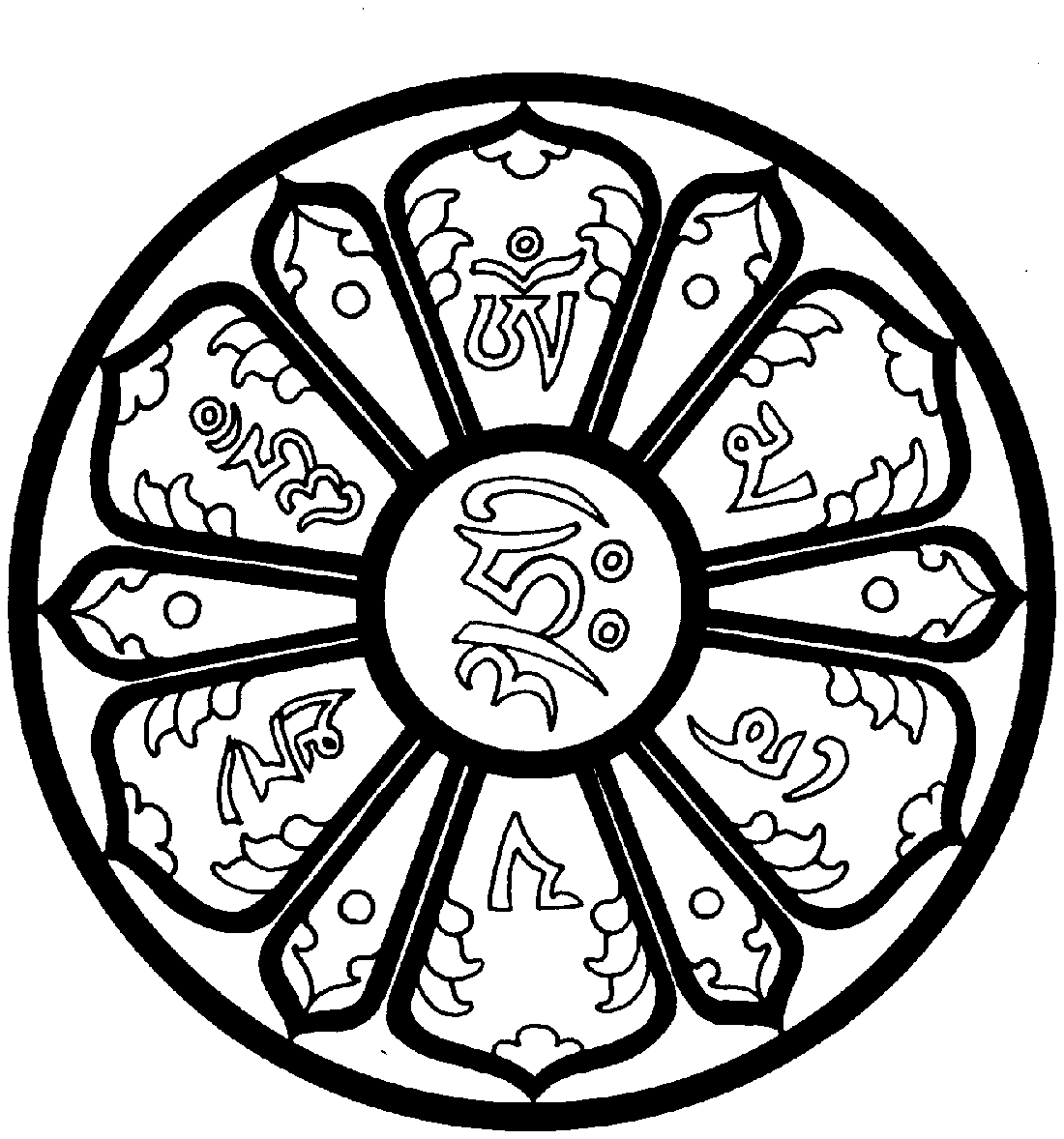,width=4cm}}}

 
\end{document}